\documentclass[prl,twocolumn,superscriptaddress,showpacs,floatfix,amsmath,amssymb,amsfonts]{revtex4}

\usepackage{graphicx} 
\usepackage{bm}

\input{Qcircuit}

\renewcommand{\Qcircuit}[1][0em]{\xymatrix @*[o] @*=<#1>}
\newcommand{\node}[2][]{{\begin{array}{c} \ _{#1}\  \\ {#2} \\ \
\end{array}}\drop\frm{o} }
\newcommand{\link}[2]{\ar @{-} [#1,#2]}

\def\bigM{\mathcal M}

\def\negspace{\!}

\def\lsub#1#2{{\vphantom{#1}}_{#2} \negspace {#1}}
\def\rsub#1#2{{#1} \negspace {\vphantom{#1}}_{#2}}

\def\ket#1{\left\lvert {#1} \right\rangle}

\def\ketsub#1#2{\rsub {\ket{#1}} {#2}}

\def\pket#1{\ketsub{#1} p}
\def\qket#1{\ketsub{#1} q}

\def\inprod#1#2{\left\langle {#1} | {#2} \right\rangle}
\def\braket{\inprod}

\def\reals{\mathbb{R}}

\def\controlled#1{\mathrm{C}_{#1}}
\def\CZ{\controlled Z}
\def\CX{\controlled X}

\def\UQ{Department of Physics, The University of Queensland, Brisbane,
Queensland 4072, Australia}

\begin{document}

\title{Universal Quantum Computation with Continuous-Variable Cluster States}

\author{Nicolas C. Menicucci}
\email{nmen@princeton.edu}
\affiliation{\UQ}
\affiliation{Department of Physics, Princeton University, Princeton, NJ 08544, USA}
\author{Peter \surname{van Loock}}
\affiliation{National Institute of Informatics, Tokyo, Japan}
\author{Mile Gu}
\affiliation{\UQ}
\author{Christian~\surname{Weedbrook}}
\affiliation{\UQ}
\author{Timothy C. Ralph}
\affiliation{\UQ}
\author{Michael A. Nielsen}
\affiliation{\UQ}

\date{\today}

\begin{abstract}
We describe a generalization of the cluster-state model of quantum
computation to continuous-variable systems, along with a proposal
for an optical implementation using squeezed-light sources, linear
optics, and homodyne detection.  For universal quantum
computation, a nonlinear element is required.  This
can be satisfied by adding to the toolbox any
single-mode non-Gaussian measurement, while the initial cluster
state itself remains Gaussian.  Homodyne detection alone suffices
to perform an arbitrary multi-mode Gaussian transformation via the
cluster state.  We also propose an experiment to demonstrate cluster-based error reduction when implementing Gaussian operations.
\end{abstract}

\pacs{03.67.Lx, 42.50.Dv}

\maketitle

%

{\bf Introduction---}One-way quantum computation~\cite{RB01a}
provides the ability to perform universal quantum computation (QC)
using only single-qubit projective measurements, given a specially
prepared and highly entangled cluster state. This is in contrast
to the traditional circuit model, where unitary evolution and
coherent control of individual qubits are required~\cite{NC00a}.
Apart from its conceptual importance, the cluster-state approach
can also lead to practical advantages. For example, the resources
required for QC using linear optics~\cite{Knill2001} can be
significantly reduced by first creating photonic cluster states
via nondeterministic gates
~\cite{Nielsen2004,Browne2005,Duan2005}. Recently, a four-qubit
cluster state has been demonstrated optically in the single-photon
regime~\cite{Walther2005}.

While qubits are typically used in QC, Lloyd and
Braunstein~\cite{LB99a} proposed the use of continuous variables for
QC and proved that only a finite set of continuous-variable (CV)
gates are needed for universal QC.  In the CV approach, the
continuous degree of freedom may be used directly, or
lower-dimensional systems may be encoded within the modes, such as
in the GKP proposal~\cite{Gottesman2001b}, which encodes one qubit
into each mode.  This allows, for instance, for the application of
standard qubit protocols to CV systems.  The optical modes of the
electromagnetic field provide an experimental testbed for these ideas~\cite{Braunstein2005a}.

In this Letter, we describe a model of universal QC using CV cluster
states.  We also propose an optical implementation of our scheme
where squeezed light sources serve as the nodes of the cluster. The main
advantage of this approach is that not only can computations with the cluster be performed deterministically, but also the preparation
of the cluster state, including connecting the nodes, can be done
unconditionally. This is in contrast to the discrete-variable
linear-optics schemes~\cite{Nielsen2004,Browne2005,Kok2005}, where
cluster states are created probabilistically. Therefore, the CV
approach appears to be particularly suited for further experimental
demonstration of the general principles of cluster-state QC.

In our optical implementation, once the cluster state has been
created, single-mode homodyne detection alone will allow for any
multi-mode Gaussian transformation to be performed on the
information contained within the cluster.  Analogously to the implementation of Clifford gates using qubit clusters, the homodyne detections
can be done in any order, a property known as {\it parallelism}. For
universal QC, in addition, only one single-mode non-Gaussian
projective measurement (e.g., photon counting) is required.
However, parallelism no longer applies to non-Gaussian measurements,
because the choice of subsequent measurement bases will
depend on the outcome of earlier measurements.  This {\it
adaptiveness} of the measurement bases is again analogous to the qubit
case when computing non-Clifford gates. While CV cluster states have been described
previously in~\cite{Zhang2006}, it is
claimed there that such states are an insufficient resource for
universal QC because of their Gaussian
character~\cite{Bartlett2002a}. In fact, they are sufficient as long
as we can perform a non-Gaussian measurement. An analogous result
holds for qubit cluster states, which can be created entirely using
Clifford group operations~\cite{Got98c} but are nevertheless
universal once a non-Clifford measurement is allowed.

Although CV cluster states can be built deterministically, it will
be impossible to create perfect CV cluster states due to the finite
degree of squeezing obtainable in the laboratory. This results in
distortions to the quantum information as it propagates through the
cluster state. We discuss these distortions (along with other
errors) and propose an experiment that demonstrates how parallelism and post-selection can be used to mitigate these effects when implementing Gaussian operations.

%

{\bf Continuous-Variable Cluster States---}Other
authors~\cite{Zhou03a,Clark05a,Hall05a} have extended the
cluster-state formalism to $d$-level systems (qudits). Here we
generalize these results to CVs.  Our use of
CVs for QC follows the standard prescription
given in~\cite{Bartlett2002a}.  The Pauli $X$ and $Z$ operators
are generalized to the Weyl-Heisenberg group, which is the group
of phase-space displacements.  For CVs, this
group is a Lie group with generators $\hat q$ and $\hat p$. These
operators satisfy the canonical commutation relation $[\hat q,
\hat p] = i$ (with $\hbar = 1$) and when exponentiated give the
finite phase-space translation operators, $X(s) = e^{-is \hat p}$ and $Z(t) = e^{it \hat q}$,
with $s,t \in \reals$.  $X(s)$ acts on a continuously indexed
computational basis state $\qket q$, an eigenstate of $\hat q$, as
$X(s) \qket q = \qket {q + s}$.  Eigenstates of $\hat p$ transform
similarly: $Z(t) \pket p = \pket {p+t}$.  
Transformation between the position and momentum basis is given by
the Fourier transform operator $F = \exp[i (\pi / 4)(\hat q^2 +
\hat p^2)]$, with $F \qket s = \pket s$.  This is the
generalization of the Hadamard gate for qubits.  The controlled
operations CNOT and CPHASE are generalized to controlled-$X$
($\CX$) and controlled-$Z$ ($\CZ$), respectively.  These operators
effect a phase-space displacement on the target by an amount
determined by the position eigenvalue of the control: $\CX = \exp(-i\hat q \otimes \hat p)$ and $\CZ = \exp(i \hat q \otimes \hat q)$,
where the order of the systems is $(\text{control} \otimes
\text{target})$.

The essence of the qubit cluster-state model of QC lies in the
one-qubit teleportation circuit~\cite{zhou-2000-62,Nielsen2005}.
This circuit gives the ability to teleport operations diagonal in
the computational basis onto the state in question {\it after} the
cluster has been prepared.  This allows dynamics to be performed
solely through measurement.  The CV analog of the one-qubit
teleportation circuit is
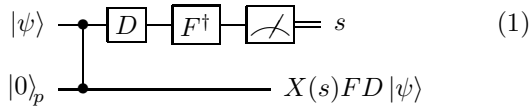
\begin{eqnarray}
\label{eq:CVwire}
    \Qcircuit @C=1em @R=1em @!R {
     \lstick{\ket \psi} & \ctrl{1}   & \gate{D} & \gate{F^\dag}
     & \meter  & \rstick{s} \cw \\
     \lstick{\pket 0}    & \control \qw & \qw & \qw
     & \rstick{X(s) F D \ket \psi} \qw
     }
\end{eqnarray}
In this diagram, $\pket 0 = (2\pi)^{-1/2} \int dq \qket q$ is a
zero-momentum eigenstate (the generalization of $\ket +$), the
controlled operation indicated is a $\CZ$ gate, and $D$ is any
operator diagonal in the computational basis (i.e., of the form
$\exp[if(\hat q)]$).  The projective measurement is of $\hat q$
and yields a real number $s$, which becomes the argument of the
displacement $X(\cdot)$ at the output of the circuit.  The
essential feature of this circuit is that the $\CZ$ gate commutes
with any diagonal operator $D$.  This means that even though $D$
is applied {\it after} the $\CZ$ gate, it acts as if it had been
applied {\it before}.  Since the operations $D$ and $F^\dag$ followed by
computational basis measurement are equivalent to a single
measurement of $D^\dag \hat p D$, manipulating quantum information
in the CV cluster is possible through projective measurements
alone.  Concatenation of these circuits makes it possible to
implement any single-mode unitary~\cite{LB99a}.

As is the case for qubits~\cite{Hein2004}, every CV cluster state
has a graph state representation, where each node in the graph is
a separate CV mode, and each link in the graph represents a $\CZ$
that has been performed between the corresponding nodes (systems).
Linear graphs, where each node has at most two links, can be used
for single-mode evolutions, but not multi-mode gates.   
The simplest
implementation of a $\CZ$ gate involves a graph state with a link
between two adjacent quantum wires:
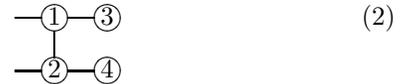
\begin{eqnarray}
\label{eq:CVCZoncluster}
    \Qcircuit[1em] @R=1em @C=1em {
    & \node{\text{1}} \link{0}{-1} \link{1}{0} & \node{\text{3}} \link{0}{-1} \\
    & \node{\text{2}} \link{0}{-1} & \node{\text{4}} \link{0}{-1}
    }
\end{eqnarray}
The lines to the left of nodes~1 and~2 indicate that a bipartite
state $\ket \psi$ will be teleported down two quantum wires to
arrive at nodes~1 and~2.  Measuring $\hat p$ on nodes~1 and~2 leaves $(X(s_1) F \otimes X(s_2) F) \CZ\ket \psi$ on nodes~3 and~4.

A small set of Hamiltonians that are polynomials in $\hat q$
(e.g., $\{\hat q, \hat q^2/2\}$), along with the Fourier transform, are sufficient to implement any single-mode Gaussian~\cite{LB99a}.  Furthermore, adding the ability to
perform a $\CZ$ operation (as described above) allows
implementation of all multi-mode Gaussians.  While this
is not sufficient for universal QC, given an encoding that maps
all qubit Clifford operations to CV Gaussian operations (the GKP
encoding being one example~\cite{Gottesman2001b}), this would be
sufficient for many quantum error correction
protocols~\cite{Got97a}.  Adding to the toolbox {\it any} single
non-Gaussian projective measurement allows for universal QC using
CV cluster states~\cite{LB99a}.

%

{\bf Optical Implementation---}Since each mode of the
electromagnetic field behaves as an independent harmonic
oscillator, we can use these modes as CV systems
for our CV cluster state.  To do this, we choose the computational
basis to be the ``position'' (amplitude) quadrature of quantum
optics for each mode.  The ``momentum'' (phase) quadrature for
each mode becomes the conjugate basis.  The commutation relations
$[a,a^\dag] = 1$ and $[\hat q,\hat p] = i$ are
satisfied by the definitions $\hat q = (a + a^\dag)/\sqrt 2$ and
$\hat p = -i(a - a^\dag)/\sqrt 2$ for each mode.  In this unitless
convention, the variance of the vacuum state (which can be
measured experimentally using homodyne detection) is given by
$\langle {\hat q^2} \rangle = \langle {\hat p^2} \rangle = 1/2$.

Construction of an ideal CV cluster state requires zero-momentum
eigenstates, which cannot be normalized and are thus unphysical.  In this optical model, they represent infinitely squeezed vacuum states, which
require infinite energy.  We can approximate them, though, by
finitely squeezed vacuum states:
\begin{eqnarray}
\label{eq:finitesqueezep}
    \pket {0,\Omega} := (\pi\Omega^2)^{-1/4} \int dp\, e^{-p^2 / 2 \Omega^2} \pket p\;,
\end{eqnarray}
with $\Omega^2 < 1$ being the variance of a Gaussian wave packet in momentum space (with $\langle {\hat p^2} \rangle = \Omega^2/2$).  The states $\qket {0,\Omega}$ are
defined analogously with $p \rightarrow q$ in
Eq.~\eqref{eq:finitesqueezep}.  Note that $\pket {0, \Omega} =
\qket {0, \Omega^{-1}}$.  The fact that these states are finitely
squeezed means that we will not have perfect fidelity while
propagating quantum information through our cluster.  This will be
addressed later.  Given the graph state that we wish to create, we
need one independently squeezed mode per node, and we need the
ability to perform a $\CZ$ gate between modes in accordance with
the graph.  This operation is a quantum nondemolition (QND)
interaction and can be implemented using two beamsplitters and two
in-line squeezers~\cite{Braunstein2005}. Alternatively, it could
be directly realized via an optical-fiber cross-Kerr
interaction~\cite{Milburn1994} in the Gaussian regime of large
photon number~\cite{Poizat1994}.  (See also Sec.~III of Ref.~\cite{Zhang2006} for further ideas.) 

Propagation down a quantum wire ($D = I$) is achieved through momentum-quadrature homodyne detection.  As discussed
previously, multi-mode Gaussian operations require only that we
can apply $D=e^{is\hat q}$ and $D=e^{it\hat q^2/2}$ for all $s, t
\in \reals$.  Applying a gate $D$ to the encoded state is achieved
by measuring the operator $D^\dag \hat p D$.  Thus, the $Z(s) =
e^{is \hat q}$ gate is implemented by measuring the operators
$Z(-s) \hat p Z(s) = \hat p - s$.  This is trivial to implement:
simply measure $\hat p$ and subtract $s$ from the result.  The
gate denoted $P(t) = \exp(it \hat q^2/2)$ is implemented by
measuring $P(-t) \hat p P(t) = \hat p + t \hat q$. Notice,
however, that by defining $\theta = \tan^{-1} (-t)$, we can
rewrite this operator as $(\hat p \cos \theta - \hat q \sin
\theta)/(\cos \theta)$, which is simply homodyne detection in a
rotated quadrature basis, followed by a rescaling of the
measurement results by a factor of $\cos \theta = (1 +
t^2)^{-1/2}$.  Thus, once the cluster has been prepared, we are
able to perform all multi-mode Gaussian operations simply through
homodyne detection.

Furthermore, analogously to implementing Clifford group operations
on qubit cluster states, all multi-mode Gaussian operations may be
implemented on CV clusters with the appropriate measurements made
{\it in any order}.  Performing the measurements in a different order
is equivalent to commuting Gaussian operations through the
(Gaussian) measurement-dependent corrections, resulting in
different corrections, but leaving the measurement bases
unchanged.  This is known as {\it parallelism} in cluster-state
QC~\cite{raussendorf-2002-49}.  Non-Gaussian operations in general
cannot be parallelized, since later measurement bases will depend
on current measurement results, a property known as {\it
adaptiveness}.

Universal QC requires the ability to implement at least one
non-Gaussian operation~\cite{LB99a}.  In our case, this will be
achieved through a measurement in a non-Gaussian basis.  While one can, in principle, use the continuous
degree of freedom directly for QC, it will almost certainly be
more practical (considering experimental errors) to encode finite
dimensional systems in the CV modes, e.g., as in the GKP
proposal~\cite{Gottesman2001b}, which encodes one qubit into each
oscillator.  In this case, the optimal non-Gaussian operation
would be tailored to implement a desirable non-Clifford unitary in
the {\it qubit space}.  Photon counting is one possibility and
fits nicely into the cluster formalism since it is already a
projective measurement.  Another option is to measure in a
nonlinear polynomial basis, such as that corresponding to the observable
$\hat p + u \hat q^2$ for any one particular choice of $u$.  This is
equivalent, in the language of Circuit~\eqref{eq:CVwire}, to
implementing the gate $D = e^{iu\hat q^3/3}$.  The GKP proposal discusses both options in more detail.
We leave the questions of encoding scheme and non-Gaussian
measurement to future work.

%

{\bf Experimental Errors---}Possible sources of experimental error
include the finite squeezing of the input states,
mixed input states (but still Gaussian), and distortions due to
the QND operation used to form the cluster.  Since {\it any}
physical implementation of our protocol will be forced to use
finitely squeezed states (because of finite energy requirements),
we will consider the effects of finite squeezing in some detail.

Finite squeezing in Eq.~\eqref{eq:finitesqueezep} modifies the output of the circuit in Circuit~\eqref{eq:CVwire}
to $\bigM X(s) F D \ket \psi$, where $\bigM$ is a distortion
that applies a Gaussian envelope in position space with
zero mean and variance $\Omega^{-2}$:
\begin{eqnarray}
\label{eq:Monpsi}
    \bigM \ket \psi \propto \int dq\, e^{q^2 \Omega^2 / 2} \qket q \lsub {\braket q \psi} q\;.
\end{eqnarray}
Notice that this is not a unitary transformation, and the state must
be renormalized after this envelope is applied.  This is also
equivalent to convolution in momentum space by a Gaussian
with variance $\Omega^2$.  Mixed input states can be accommodated in
this analysis (in the Wigner representation) simply by allowing the convolution width to be
independent of the width of the Gaussian envelope. Thus, the
transformation implemented by each measurement, which used to
consist solely of $F$s, $D$s, and phase-space displacements, now
includes a ubiquitous distortion at each step in the evolution.  The
severity of this distortion depends inversely on the amount by which
the sources are squeezed.

Concatenated circuits of the form~\eqref{eq:CVwire} apply the
transformation $\dotsm \bigM X(s_2) F D_2 \bigM X(s_1) F D_1 \ket
\psi$ to the input.    
Alternatively, we can gather the fixed distortions to one end of the operation and transform this into the useful form $U_0(s_1,\dotsc,s_n)\widetilde \bigM (s_1,\dotsc,s_n) \ket \psi$, where $U_0$ is the unitary that would be applied in the case of an ideal cluster, and $\widetilde \bigM (s_1,\dotsc,s_n) \ket \psi$ is a distorted initial state, with the distortion now depending on both the measurement results and the gate to be implemented.  The effect is the same
for multi-mode gates: at each measurement step, a fixed distortion $\bigM$
is applied to each mode.  Specifically, in the case of the
$\CZ$ gate, the resulting output is $\bigl( \bigM X(s_1) F \otimes
\bigM X(s_2) F \bigr) C_Z \ket \psi$.  
The distortion operations in the multi-mode case can similarly be gathered to the right while becoming measurement- and gate-dependent.  Errors in the QND operation can be modeled as additive
Gaussian noise, which has a similar distorting effect, the strength
of which scales as the number of links in the cluster's graph.

%

{\bf Experimental Proposal for Cluster-Based Error Reduction---}
Parallelism, which is a feature particular to cluster-state QC, along with post-selection, can be used to reduce the impact of the errors described in the last section when implementing Gaussian
operations.  We propose an experiment to demonstrate this.  For concreteness, consider a linear cluster of five nodes (although any number greater than three will suffice):
\begin{eqnarray}
\label{eq:postselectsqueeze}
	\Qcircuit[1em] @R=1em @C=1em {
	& \node{1} \link{0}{-1} & \node{2} \link{0}{-1} & \node{3} \link{0}{-1} & \node{4} \link{0}{-1} & \node{5} \link{0}{-1}
	}
\end{eqnarray}
(The line to the left of the first node indicates where this cluster might be attached to another one.)  Many simple Gaussian operations may be implemented on this cluster through homodyne detection on the first four nodes.  With each such measurement there is the possibility of the resulting distortion severely affecting the quantum state in a measurement-dependent way (see the previous section).  However, if we choose to delay applying the QND operation that connects node~1 to node~2, we can isolate nodes 2--5 into a ``mini-cluster,'' which is separate from the quantum state to be acted upon.  By measuring nodes~3 and~4 {\it before} attaching the mini-cluster to the input state, we can calculate the effect of the distortion from these two nodes before that distortion ever affects the state.  If this distortion does not preserve the Wigner phase-space region likely to
be occupied by the input state (which depends on the chosen
encoding), we discard this mini-cluster and try again. If it does,
we perform the QND operation to attach nodes~1 and~2.  We now have only two ``dangerous'' measurements to make (on the newly attached nodes) instead of four, with the output appearing
on node~5.  State tomography can be used to compare the fidelity of these two approaches.

This technique generalizes easily to multi-qubit operations and can, in fact, be applied to mini-clusters
implementing {\it any} Gaussian operation.  The greatest
benefit will be for Gaussians that require many measurements.
While we have ``bent the rules'' of cluster-state QC a bit by
delaying attachment of the mini-cluster and by
post-selecting mini-clusters based on measurement results, this
may yet prove to be a practical procedure for dealing with
experimental errors.  This result has the flavor of Ref.~\cite{Dawson2006}, wherein it is shown that through post-selection the reliability of an error-correcting ancilla cluster (called a ``telecorrector'') can be guaranteed before it is attached to the state to be corrected.

%

{\bf Conclusion---}We have generalized the notion of universal
cluster-state quantum computation to continuous-variable systems. We have
proposed an optical implementation that uses squeezed light sources
and quantum nondemolition operations to build a Gaussian cluster
state.  Homodyne detection alone
suffices to implement all multi-mode Gaussian operations using the cluster state, with the addition of one non-Gaussian measurement allowing for universal quantum computation. Many of
the properties of qubit-cluster computation also apply to the
continuous-variable case, including parallelism and adaptiveness.
Within the continuous-variable approach, a lower-dimensional
encoding scheme will most likely be required for experimental viability. Due to their Gaussian
nature and deterministic method of construction, we expect that continuous-variable
cluster states will allow for further experimental demonstrations of
the principles of cluster-state quantum computation.  We have proposed such an experiment to demonstrate improvement in the fidelity of Gaussian operations using post-selection and parallelism.

%

{\bf Acknowledgments---}We thank Gerard Milburn for help with the
details of the optical implementation, as well as Mark de Burgh, Mark Dowling, Henry
Haselgrove, and Kae Nemoto for
useful discussions.  We also thank the Australian Research Council for their
support of this work.  NCM was supported by a United States NDSEG
Fellowship, and PvL acknowledges funding from MIC in Japan.


\bibliography{CVCS_Letter}

\begin{thebibliography}{26}
\expandafter\ifx\csname natexlab\endcsname\relax\def\natexlab#1{#1}\fi
\expandafter\ifx\csname bibnamefont\endcsname\relax
  \def\bibnamefont#1{#1}\fi
\expandafter\ifx\csname bibfnamefont\endcsname\relax
  \def\bibfnamefont#1{#1}\fi
\expandafter\ifx\csname citenamefont\endcsname\relax
  \def\citenamefont#1{#1}\fi
\expandafter\ifx\csname url\endcsname\relax
  \def\url#1{\texttt{#1}}\fi
\expandafter\ifx\csname urlprefix\endcsname\relax\def\urlprefix{URL }\fi
\providecommand{\bibinfo}[2]{#2}
\providecommand{\eprint}[2][]{\url{#2}}

\bibitem[{\citenamefont{Raussendorf and Briegel}(2001)}]{RB01a}
\bibinfo{author}{\bibfnamefont{R.}~\bibnamefont{Raussendorf}} \bibnamefont{and}
  \bibinfo{author}{\bibfnamefont{H.~J.} \bibnamefont{Briegel}},
  \bibinfo{journal}{Phys. Rev. Lett.} \textbf{\bibinfo{volume}{86}},
  \bibinfo{pages}{5188} (\bibinfo{year}{2001}).

\bibitem[{\citenamefont{Nielsen and Chuang}(2000)}]{NC00a}
\bibinfo{author}{\bibfnamefont{M.~A.} \bibnamefont{Nielsen}} \bibnamefont{and}
  \bibinfo{author}{\bibfnamefont{I.~L.} \bibnamefont{Chuang}},
  \emph{\bibinfo{title}{Quantum Computation and Quantum Information}}
  (\bibinfo{publisher}{Cambridge University Press},
  \bibinfo{address}{Cambridge, UK}, \bibinfo{year}{2000}).

\bibitem[{\citenamefont{Knill et~al.}(2001)\citenamefont{Knill, Laflamme, and
  Milburn}}]{Knill2001}
\bibinfo{author}{\bibfnamefont{E.}~\bibnamefont{Knill}},
  \bibinfo{author}{\bibfnamefont{R.}~\bibnamefont{Laflamme}}, \bibnamefont{and}
  \bibinfo{author}{\bibfnamefont{G.~J.} \bibnamefont{Milburn}},
  \bibinfo{journal}{Nature} \textbf{\bibinfo{volume}{409}}, \bibinfo{pages}{46}
  (\bibinfo{year}{2001}).

\bibitem[{\citenamefont{Nielsen}(2004)}]{Nielsen2004}
\bibinfo{author}{\bibfnamefont{M.~A.} \bibnamefont{Nielsen}},
  \bibinfo{journal}{Phys. Rev. Lett.} \textbf{\bibinfo{volume}{93}},
  \bibinfo{eid}{040503} (\bibinfo{year}{2004}).

\bibitem[{\citenamefont{Duan and Raussendorf}(2005)}]{Duan2005}
\bibinfo{author}{\bibfnamefont{L.-M.} \bibnamefont{Duan}} \bibnamefont{and}
  \bibinfo{author}{\bibfnamefont{R.}~\bibnamefont{Raussendorf}},
  \bibinfo{journal}{Phys. Rev. Lett.} \textbf{\bibinfo{volume}{95}},
  \bibinfo{eid}{080503} (\bibinfo{year}{2005}).

\bibitem[{\citenamefont{Browne and Rudolph}(2005)}]{Browne2005}
\bibinfo{author}{\bibfnamefont{D.~E.} \bibnamefont{Browne}} \bibnamefont{and}
  \bibinfo{author}{\bibfnamefont{T.}~\bibnamefont{Rudolph}},
  \bibinfo{journal}{Phys. Rev. Lett.} \textbf{\bibinfo{volume}{95}},
  \bibinfo{pages}{010501} (\bibinfo{year}{2005}).

\bibitem[{\citenamefont{Walther et~al.}(2005)\citenamefont{Walther, Resch,
  Rudolph, Schenck, Weinfurter, Vedral, Aspelmeyer, and
  Zeilinger}}]{Walther2005}
\bibinfo{author}{\bibfnamefont{P.}~\bibnamefont{Walther}},
  \bibinfo{author}{\bibfnamefont{K.~J.} \bibnamefont{Resch}},
  \bibinfo{author}{\bibfnamefont{T.}~\bibnamefont{Rudolph}},
  \bibinfo{author}{\bibfnamefont{E.}~\bibnamefont{Schenck}},
  \bibinfo{author}{\bibfnamefont{H.}~\bibnamefont{Weinfurter}},
  \bibinfo{author}{\bibfnamefont{V.}~\bibnamefont{Vedral}},
  \bibinfo{author}{\bibfnamefont{M.}~\bibnamefont{Aspelmeyer}},
  \bibnamefont{and}
  \bibinfo{author}{\bibfnamefont{A.}~\bibnamefont{Zeilinger}},
  \bibinfo{journal}{Nature} \textbf{\bibinfo{volume}{434}},
  \bibinfo{pages}{169} (\bibinfo{year}{2005}).

\bibitem[{\citenamefont{Lloyd and Braunstein}(1999)}]{LB99a}
\bibinfo{author}{\bibfnamefont{S.}~\bibnamefont{Lloyd}} \bibnamefont{and}
  \bibinfo{author}{\bibfnamefont{S.~L.} \bibnamefont{Braunstein}},
  \bibinfo{journal}{Phys. Rev. Lett.} \textbf{\bibinfo{volume}{82}},
  \bibinfo{pages}{1784} (\bibinfo{year}{1999}).

\bibitem[{\citenamefont{Gottesman et~al.}(2001)\citenamefont{Gottesman, Kitaev,
  and Preskill}}]{Gottesman2001b}
\bibinfo{author}{\bibfnamefont{D.}~\bibnamefont{Gottesman}},
  \bibinfo{author}{\bibfnamefont{A.}~\bibnamefont{Kitaev}}, \bibnamefont{and}
  \bibinfo{author}{\bibfnamefont{J.}~\bibnamefont{Preskill}},
  \bibinfo{journal}{Phys. Rev. A} \textbf{\bibinfo{volume}{64}},
  \bibinfo{pages}{012310} (\bibinfo{year}{2001}).

\bibitem[{\citenamefont{Braunstein and van Loock}(2005)}]{Braunstein2005a}
\bibinfo{author}{\bibfnamefont{S.~L.} \bibnamefont{Braunstein}}
  \bibnamefont{and} \bibinfo{author}{\bibfnamefont{P.}~\bibnamefont{van
  Loock}}, \bibinfo{journal}{Rev. Mod. Phys.} \textbf{\bibinfo{volume}{77}},
  \bibinfo{eid}{513} (\bibinfo{year}{2005}).

\bibitem[{\citenamefont{Kok et~al.}(2005)\citenamefont{Kok, Munro, Nemoto,
  Ralph, Dowling, and Milburn}}]{Kok2005}
\bibinfo{author}{\bibfnamefont{P.}~\bibnamefont{Kok}},
  \bibinfo{author}{\bibfnamefont{W.~J.} \bibnamefont{Munro}},
  \bibinfo{author}{\bibfnamefont{K.}~\bibnamefont{Nemoto}},
  \bibinfo{author}{\bibfnamefont{T.~C.} \bibnamefont{Ralph}},
  \bibinfo{author}{\bibfnamefont{J.~P.} \bibnamefont{Dowling}},
  \bibnamefont{and} \bibinfo{author}{\bibfnamefont{G.~J.}
  \bibnamefont{Milburn}} (\bibinfo{year}{2005}), \eprint{quant-ph/0512071}.

\bibitem[{\citenamefont{Zhang and Braunstein}(2006)}]{Zhang2006}
\bibinfo{author}{\bibfnamefont{J.}~\bibnamefont{Zhang}} \bibnamefont{and}
  \bibinfo{author}{\bibfnamefont{S.~L.} \bibnamefont{Braunstein}},
  \bibinfo{journal}{Phys. Rev. A} \textbf{\bibinfo{volume}{73}},
  \bibinfo{eid}{032318} (\bibinfo{year}{2006}).

\bibitem[{\citenamefont{Bartlett et~al.}(2002)\citenamefont{Bartlett, Sanders,
  Braunstein, and Nemoto}}]{Bartlett2002a}
\bibinfo{author}{\bibfnamefont{S.~D.} \bibnamefont{Bartlett}},
  \bibinfo{author}{\bibfnamefont{B.~C.} \bibnamefont{Sanders}},
  \bibinfo{author}{\bibfnamefont{S.~L.} \bibnamefont{Braunstein}},
  \bibnamefont{and} \bibinfo{author}{\bibfnamefont{K.}~\bibnamefont{Nemoto}},
  \bibinfo{journal}{Phys. Rev. Lett.} \textbf{\bibinfo{volume}{88}},
  \bibinfo{pages}{097904} (\bibinfo{year}{2002}).

\bibitem[{\citenamefont{Gottesman}(1998)}]{Got98c}
\bibinfo{author}{\bibfnamefont{D.}~\bibnamefont{Gottesman}}
  (\bibinfo{year}{1998}), \eprint{quant-ph/9807006}.

\bibitem[{\citenamefont{Zhou et~al.}(2003)\citenamefont{Zhou, Zeng, Xu, and
  Sun}}]{Zhou03a}
\bibinfo{author}{\bibfnamefont{D.~L.} \bibnamefont{Zhou}},
  \bibinfo{author}{\bibfnamefont{B.}~\bibnamefont{Zeng}},
  \bibinfo{author}{\bibfnamefont{Z.}~\bibnamefont{Xu}}, \bibnamefont{and}
  \bibinfo{author}{\bibfnamefont{C.~P.} \bibnamefont{Sun}},
  \bibinfo{journal}{Phys. Rev. A} \textbf{\bibinfo{volume}{68}},
  \bibinfo{pages}{062303} (\bibinfo{year}{2003}).

\bibitem[{\citenamefont{Clark}(2006)}]{Clark05a}
\bibinfo{author}{\bibfnamefont{S.}~\bibnamefont{Clark}}, \bibinfo{journal}{J.
  Phys. A: Math. Gen} \textbf{\bibinfo{volume}{39}}, \bibinfo{pages}{2701}
  (\bibinfo{year}{2006}).

\bibitem[{\citenamefont{Hall}(2005)}]{Hall05a}
\bibinfo{author}{\bibfnamefont{W.}~\bibnamefont{Hall}} (\bibinfo{year}{2005}),
  \eprint{quant-ph/0512130}.

\bibitem[{\citenamefont{Zhou et~al.}(2000)\citenamefont{Zhou, Leung, and
  Chuang}}]{zhou-2000-62}
\bibinfo{author}{\bibfnamefont{X.}~\bibnamefont{Zhou}},
  \bibinfo{author}{\bibfnamefont{D.~W.} \bibnamefont{Leung}}, \bibnamefont{and}
  \bibinfo{author}{\bibfnamefont{I.~L.} \bibnamefont{Chuang}},
  \bibinfo{journal}{Phys. Rev. A} \textbf{\bibinfo{volume}{62}},
  \bibinfo{pages}{052316} (\bibinfo{year}{2000}).

\bibitem[{\citenamefont{Nielsen}(2006)}]{Nielsen2005}
\bibinfo{author}{\bibfnamefont{M.~A.} \bibnamefont{Nielsen}},
  \bibinfo{journal}{Rep. Math. Phys.} \textbf{\bibinfo{volume}{57}},
  \bibinfo{pages}{147} (\bibinfo{year}{2006}).

\bibitem[{\citenamefont{Hein et~al.}(2004)\citenamefont{Hein, Eisert, and
  Briegel}}]{Hein2004}
\bibinfo{author}{\bibfnamefont{M.}~\bibnamefont{Hein}},
  \bibinfo{author}{\bibfnamefont{J.}~\bibnamefont{Eisert}}, \bibnamefont{and}
  \bibinfo{author}{\bibfnamefont{H.~J.} \bibnamefont{Briegel}},
  \bibinfo{journal}{Phys. Rev. A} \textbf{\bibinfo{volume}{69}},
  \bibinfo{eid}{062311} (\bibinfo{year}{2004}).

\bibitem[{\citenamefont{Gottesman}(1997)}]{Got97a}
\bibinfo{author}{\bibfnamefont{D.}~\bibnamefont{Gottesman}}, Ph.D. thesis,
  \bibinfo{school}{California Institute of Technology},
  \bibinfo{address}{Pasadena, CA} (\bibinfo{year}{1997}),
  \eprint{quant-ph/9705052}.

\bibitem[{\citenamefont{Braunstein}(2005)}]{Braunstein2005}
\bibinfo{author}{\bibfnamefont{S.~L.} \bibnamefont{Braunstein}},
  \bibinfo{journal}{Phys. Rev. A} \textbf{\bibinfo{volume}{71}},
  \bibinfo{eid}{055801} (\bibinfo{year}{2005}).

\bibitem[{\citenamefont{Walls and Milburn}(1994)}]{Milburn1994}
\bibinfo{author}{\bibfnamefont{D.~F.} \bibnamefont{Walls}} \bibnamefont{and}
  \bibinfo{author}{\bibfnamefont{G.~J.} \bibnamefont{Milburn}},
  \emph{\bibinfo{title}{Quantum Optics}} (\bibinfo{publisher}{Springer,
  Berlin}, \bibinfo{year}{1994}).

\bibitem[{\citenamefont{Poizat et~al.}(1994)\citenamefont{Poizat, Roch, and
  Grangier}}]{Poizat1994}
\bibinfo{author}{\bibfnamefont{J.-P.} \bibnamefont{Poizat}},
  \bibinfo{author}{\bibfnamefont{J.~F.} \bibnamefont{Roch}}, \bibnamefont{and}
  \bibinfo{author}{\bibfnamefont{P.}~\bibnamefont{Grangier}},
  \bibinfo{journal}{Ann. Phys. (Paris)} \textbf{\bibinfo{volume}{19}},
  \bibinfo{pages}{265} (\bibinfo{year}{1994}).

\bibitem[{\citenamefont{Raussendorf et~al.}(2002)\citenamefont{Raussendorf,
  Browne, and Briegel}}]{raussendorf-2002-49}
\bibinfo{author}{\bibfnamefont{R.}~\bibnamefont{Raussendorf}},
  \bibinfo{author}{\bibfnamefont{D.~E.} \bibnamefont{Browne}},
  \bibnamefont{and} \bibinfo{author}{\bibfnamefont{H.~J.}
  \bibnamefont{Briegel}}, \bibinfo{journal}{J. Mod. Opt.}
  \textbf{\bibinfo{volume}{49}}, \bibinfo{pages}{1299} (\bibinfo{year}{2002}).

\bibitem[{\citenamefont{Dawson et~al.}(2006)\citenamefont{Dawson, Haselgrove,
  and Nielsen}}]{Dawson2006}
\bibinfo{author}{\bibfnamefont{C.~M.} \bibnamefont{Dawson}},
  \bibinfo{author}{\bibfnamefont{H.~L.} \bibnamefont{Haselgrove}},
  \bibnamefont{and} \bibinfo{author}{\bibfnamefont{M.~A.}
  \bibnamefont{Nielsen}}, \bibinfo{journal}{Phys. Rev. Lett.}
  \textbf{\bibinfo{volume}{96}}, \bibinfo{eid}{020501}
(\bibinfo{year}{2006}).

\end{thebibliography}

\end{document}